# Anyonic Braiding in Optical Lattices

Chuanwei Zhang, V.W. Scarola, Sumanta Tewari and S. Das Sarma

Condensed Matter Theory Center, Department of Physics, University of Maryland, College Park, 20742, USA

## Abstract

Topological quantum states of matter, both Abelian and non-Abelian, are characterized by excitations whose wavefunctions undergo non-trivial statistical transformations as one excitation is moved (braided) around another. Topological quantum computation proposes to use the topological protection and the braiding statistics of a non-Abelian topological state to perform quantum computation. The enormous technological prospect of topological quantum computation provides new motivation for experimentally observing a topological state. Here we explicitly work out a realistic experimental scheme to create and braid the Abelian topological excitations in the Kitaev model built on a tunable robust system, a cold atom optical lattice. We also demonstrate how to detect the key feature of these excitations, their braiding statistics. Observation of this statistics would directly establish the existence of anyons, quantum particles which are neither fermions nor bosons. In addition to establishing topological matter, the experimental scheme we develop here can also be adapted to a non-Abelian topological state, supported by the same Kitaev model but in a different parameter regime, to eventually build topologically protected quantum gates.

Quantum computers utilize intrinsically quantum mechanical properties of matter to perform some difficult computational tasks, such as prime factorization, exponentially faster than classical computers[1]. However, quantum computation, while being possible in principle, is turning out to be difficult because quantum error corrections are very hard to carry out, and without error correction, no substantial computation process, quantum or classical, is feasible. Unfortunately, the tolerance for errors in a quantum error correction scheme [2] is very small, which leads to the necessity for a very large number of additional 'physical' qubits (quantum bits) for each 'logical' qubit in a complex quantum computer architecture. In this context, a revolutionary recent development is the concept of topological quantum computation[3-8]. A topological quantum computer is robustly protected from local errors by the physical hardware and one does not, in principle, need any software-level quantum error correction protocols that are required for a regular qubit-based quantum computer[9-14].

The topological state of matter has enhanced ground state symmetries which do not exist in the bare Hamiltonian of the system. This enhanced topological symmetry protects the ground state from quantum errors associated with external fluctuations providing the robustness needed for fault-tolerant quantum computation.

The early proposal [3,4] for topological quantum computation was studied mostly as a deep mathematical curiosity because no physical implementation was thought to be possible. This all changed recently when serious specific suggestions [15] were made to study non-Abelian topological order through manipulating delicate fractional Quantum Hall (FQH) states in low-temperature two-dimensional electron layers as an initial step to building a topological quantum computer in the laboratory. These suggestions have generated a great deal of interest in a broad spectrum of disciplines including physics, mathematics, computer science, and of course, quantum computation. Several groups are currently working on carrying out experiments to see if FQH topological quantum computation is feasible even as a matter of principle.

The main problem in carrying out topological quantum computation using FQH states is that there is essentially no experimental evidence determining whether the actual experimentally observed 5/2 and 12/5 FQH states are in fact non-Abelian states, allowing quantum computation. Therefore, initial experimental work will be directed entirely toward an experimental demonstration of the topological nature of these states. Such an experimental demonstration by itself will be important since topological quantum states have never been directly observed experimentally.

In this article, we discuss a different situation, where the topological nature of the quantum state is assured by design, i.e. the quantum state is constructed as a topological state. These are model systems controlled by Hamiltonians whose properties guarantee topological protection. The most famous example of this is the magnetic Kitaev lattice, described in the pioneering papers [2,3] on topological quantum computation. The Kitaev model is an exactly soluble lattice model that carries excitations with both Abelian and non-Abelian anyonic braiding statistics, which are the hallmarks of topological quantum matter, i.e. excitations which do not obey ordinary bosonic and fermionic statistics, but are anyons with more complex statistical behavior [5] arising from braiding. The usual definition of permutation statistics for fermions and bosons can be thought of as a half braid of one particle around another of the same species followed by a translation to effectively exchange the positions of the two particles. The net result is an overall gain in a plus or minus sign in the wavefunction for bosons or fermions, respectively. Note that a full braid (a closed loop) does not result in a sign change. The Abelian anyon wavefunction, by contrast, acquires a phase factor upon a full braid of one anyon around another while a braid of non-Abelian anyons unitarily transforms the wavefunction as a vector in a finite-dimensional Hilbert space[5], making the successive braiding operations non-commutative.

Precise proposals to construct an artificial Kitaev lattice using atomic optical lattices have recently been made in the literature[16,17]. So we know how to make a Kitaev lattice, and we also know that such a lattice supports both Abelian and non-Abelian topological phases, and, in both phases, the topological robustness is guaranteed. In addition, recent numerical results[18] show that weak, local perturbations (e.g. a stray Zeeman field or unwanted interaction terms) do not destroy topological order. But the problem that has remained unclear, and what we discuss in this article, is a way to carry out the topological gating operations, called 'braiding' in the technical literature, on such an optical lattice based topological system and subsequently detect the results. Our suggested braiding technique, which requires successive manipulations of adjacent lattice sites that we work out in detail, can not only be implemented on the proposed Kitaev optical lattice, but can also be used in other proposals for doing topological quantum computation in optical lattices; a bosonic model involving the extended Hubbard model[19] is one example.

We note that a proposal[20] for observing the Abelian anyonic phase in a rotating BEC consisting of a small number of atoms has recently been made in the literature. The proposed system is essentially a small continuous quantum Hall liquid and completely different from the large discrete Kitaev lattice system discussed here. As we will see the origin and creation of anyonic excitations, the braiding operation, the detection of statistics, and even the large size ($\sim 10^5$ atoms) of the lattice system are completely different in ways which make optical lattices, and specifically the Kitaev model, a more attractive candidate for realizing and detecting topological matter.

We stress that the techniques for braiding and read-out proposed here provide a necessary first step in eventually performing topological quantum computation in optical lattices. Here we should make a clear distinction between quantum computation using Abelian and non-Abelian systems. An Abelian anyonic system has two degenerate ground states which cannot mix by a weak local external perturbation in the sense that the errors induced by local perturbations are exponentially suppressed $\sim \exp(-L/\xi)$, where $L$ is the linear size of the system, and $\xi$ is a characteristic length inversely proportional to the excitation gap [3,4]. In the ground state sector, one thus has a topologically protected two-state system which, on multiply connected surfaces, can be duplicated to produce an array of qubits and used for topological quantum memory [3,4]. Quantum computation can then be accomplished by devising the conventional *non-topologically protected* single- and two-qubit gates. In a non-Abelian topological phase (qubits are topologically protected here as well), on the other hand, the quantum gates can be constructed simply by braiding one quasiparticle around another thereby exploiting the statistical effects of these braids. Therefore, implementation of these gates is immune to local deformations of the braiding trajectory since the effects of the braid transformations are statistical and hence only depend on the braid topologies. In this sense the putative quantum gates are noiseless.

Recently Ioffe *et al.* [21] proposed building Josephson-junction arrays to simulate the quantum dimer model on some frustrated lattices which in turn supports topological phases and quantum computation in the Abelian setting. However, the corresponding Josephson-junction architecture for a non-Abelian phase is extremely complex [22]. The beauty of the Kitaev model is that, in contrast to the quantum dimer model, it can support both the Abelian and the non-Abelian phases just by varying the optical lattice parameters. Optical lattices offer a much more coherent and tunable quantum system than the Josephson-junction system necessary for the implementation of the topological phases. Therefore, with a view to an eventual topological quantum computer built with the non-Abelian phase, we focus our attention here on the Kitaev optical lattice model. Our work here clarifies the nature of the elementary excitations, the origin of the topological phase change acquired by the wavefunction upon braiding, and how one can experimentally carry out the braiding operation and detect the braiding statistics in the Abelian phase of the Kitaev lattice, all of which are directly applicable to the more complex non-Abelian phase. In the non-Abelian phase although the precise mathematical construction of the braiding operator remains, as of now, unknown [23] and work in this direction is in progress, it is clear that on the operational level it involves the same successive single site spin manipulations as we discuss here, and so the underlying experimental techniques remain the same. Thus, we take an important first step towards topological quantum computation in optical lattices. Furthermore, even the simple observation of Abelian topological ("anyonic") properties in an optical lattice along the lines of our proposed braiding procedure and the subsequent read-out scheme will be a breakthrough achievement in itself, since anyonic statistics have never been directly demonstrated in any experimental system.

The Kitaev model describes a set of individual spins placed at the vertices of a two dimensional honeycomb lattice with a spatially anisotropic interaction between neighboring spins. The Hamiltonian is given by [4]:

$$H = -J_x \sum_{x-link} \sigma_j^x \sigma_k^x - J_y \sum_{y-link} \sigma_j^y \sigma_k^y - J_z \sum_{z-link} \sigma_j^z \sigma_k^z , \quad (1)$$

where $J_\alpha$ are interaction parameters and $\sigma_j^\alpha$ are the Pauli matrices at the site $j$, for $\alpha = x, y, z$. Normally, neighboring spins in Heisenberg models interact isotropically so that the spin-spin interaction does not depend on the spatial direction between neighbors. In the above model, however, neighboring spins along links pointing in different directions (see Fig. 1**a**) interact differently. This model contains conserved quantities allowing an exact solution for both the ground and excited states. Two distinct regimes, defined solely by the interaction parameters, carry excitations with either Abelian or non-Abelian braiding statistics.

Ultra-cold atoms in optical lattices offer the possibility of designing such anisotropic lattice models[16,17]. Without loss of generality we focus on the proposal in Ref. 16 and present a modified implementation scheme for $^{87}$Rb atoms with a slightly different laser configuration. Consider a $^{87}$Rb Bose-Einstein condensate prepared in the hyperfine ground state $|\downarrow\rangle \equiv |F = 1, m_F = -1\rangle$ and confined to a honeycomb optical lattice in a single two dimensional ($XY$) plane, where $F$ and $m_F$ denote the total angular momentum and the magnetic quantum number of the hyperfine state. The atomic dynamics along the $\vec{Z}$ axis are frozen out by optical traps with a high trapping frequency [24]. Two hyperfine ground states $|\uparrow\rangle = |F = 2, m_F = -2\rangle$ and $|\downarrow\rangle \equiv |F = 1, m_F = -1\rangle$ are defined as the effective atomic spin. We apply three pairs of far red-detuned interfering traveling laser beams (wavelength $\lambda_0 = 850$nm) above the $XY$-plane with an angle $\varphi_{//} = 2\arcsin(\lambda_0 / \lambda_s \sqrt{3})$, where $\lambda_s = 787.6$nm is the wavelength of the spin-dependent laser beams described below. The projections on the *XY*-plane of the three pairs of lasers are along the angles $\pm \pi/6$ and $\pi/2$ respectively. These interfering laser beams form a traveling wave along the $\vec{Z}$ direction, but a spin-independent honeycomb optical lattice structure with the lattice spacing $a = \lambda_s/\sqrt{3}$ in the *XY* plane. The potential barrier between neighboring atoms in the honeycomb lattice is adiabatically ramped up to about $V_0 = 14 E_R$ to obtain a Mott insulator state with one atom per lattice site [25,26], where $E_R = h^2/2m\lambda_0^2$ is the recoil energy for Rb atoms.

In this honeycomb lattice, we can engineer the anisotropic spin-spin interactions $J_\nu \sigma_i^\nu \sigma_j^\nu$ in Equation (1) using additional spin-dependent standing wave laser beams in the *XY*-plane. With properly chosen laser configurations [16], a spin dependent potential $V_\sigma^\nu = V_+^\nu |+\rangle_\nu \langle+| + V_-^\nu |-\rangle_\nu \langle-|$ (the spatially varying parts are omitted here) along different tunneling directions $\nu = x, y, z$ can be generated, where $|\pm\rangle_\nu$ are the eigenstates of the corresponding Pauli operator $\sigma^\nu$. We adjust $V_+^\nu$ and $V_-^\nu$ so that atoms can tunnel with a rate $t_{+\nu}$ only when it is in the eigenstate $|+\rangle_\nu$, which yield the effective spin-spin exchange interaction $J_\nu \sigma_i^\nu \sigma_j^\nu$ with the interaction strength $J_\nu = -t_{+\nu}^2/U$. Here $U$ is the on-site interaction energy of atoms.

For simplicity, in the following we show how to generate the spin-spin interaction $J_z \sigma_i^z \sigma_j^z$ in the Hamiltonian (1) as an example while other spin-spin interaction terms can be created using a similar procedure [16]. The standing wave laser beam used for generating spin-dependent tunneling is along the *z*-link direction and has a detuning $\Delta_0 \approx -2\pi \times 3600 \text{GHz}$ to the $5^2 P_{3/2}$ state (corresponding to a wavelength $\lambda_s = 787.6 \text{nm}$). This laser beam forms a blue-detuning potential for atoms with spin $|\uparrow\rangle$, but a red-detuning potential for $|\downarrow\rangle$ atoms. For instance, with a properly chosen laser intensity, the spin-dependent potential barrier may be set as $\overline{V}_\downarrow = 8 E_R$ and $\overline{V}_\uparrow = -4 E_R$, which, combined with the spin-independent lattice potential barrier $V_0 = 14 E_R$, yield the total effective spin-dependent lattice potential barrier $\widetilde{V}_\downarrow = 22 E_R$ and $\widetilde{V}_\uparrow = 10 E_R$ for neighboring atoms in the honeycomb lattice. Therefore the tunneling rates for two spin states satisfy $t_\uparrow/t_\downarrow \gg 1$, which, as shown in Ref. [16], leads to the spin-spin interaction $J_z \sigma_i^z \sigma_j^z$ with $J_z \sim t_\uparrow^2/U$ [16]. For $^{87}$Rb atoms, we estimate the time scale for the spin-spin interaction $h/J_z \sim 10\,\text{ms}$. The spin-dependent lattice is adiabatically ramped up and atoms in the optical lattices follow the time-varying Hamiltonian and reach the final ground state $|\psi_g\rangle$ of the Kitaev model, which provides the starting point to our analysis. By carefully tuning the spin-dependent lattice depth in different directions, one can in principle access all phases of the Kitaev model.

We briefly discuss two technical issues with this scheme: spontaneous emission and finite temperatures. Because of the large detuning of the spin-dependent lasers, the spontaneous emission rate for atoms is suppressed and may be estimated $2\Gamma(|\bar{V}_\uparrow|)/\hbar|\Delta_0| \sim 0.3\,\text{s}^{-1}$, where $\Gamma \approx 2\pi \times 6\,\text{MHz}$ is the decay rate of the excited hyperfine state. This decay rate is sufficient to allow the preparation of the initial ground states as well as many spin operations. In addition, the temperature of the system needs to be much lower than the spin-spin interaction strength $T \ll J_z/k_B \sim 1\,\text{nK}$, which sets a strict requirement for experiments. Larger temperatures will populate the system with an excess of unwanted excitations.

Given the ability to engineer the ground state of the above model, how do we create excitations? In what follows we consider the limit defining the Abelian phase, $J_z \gg J_x, J_y$, as a conceptual first step toward realizing non-trivial braiding statistics. In the case $J_x = J_y = 0$, the low energy Hilbert space is spanned by aligned pairs of $z$-links ($\uparrow\uparrow$ or $\downarrow\downarrow$) on neighboring sites. The direction of alignment (up or down), however, is not fixed energetically. The ground state, therefore, is highly degenerate. Doing degenerate perturbation theory in $J_x$ and $J_y$, while preserving the ground state subspace, the original Hamiltonian reduces to[3]: $H_{\text{eff}} = -J_{\text{eff}} \sum_p W_p$, where $J_{\text{eff}} = J_x^2 J_y^2 / 16|J_z|^3$ and the sum is over all plaquettes (hexagons). $H_{\text{eff}}$ is unitarily equivalent to the toric code[3], in the terminology of topological quantum computation. It is written in terms of the operator associated with lattice plaquettes, $W_p = \sigma_1^x \sigma_2^y \sigma_3^z \sigma_4^x \sigma_5^y \sigma_6^z$, (see Fig. 1**a**), which can have eigenvalues $+1$ or $-1$. $W_p$ tests the spin orientation around hexagons. The ground state is defined as a superposition of all spin configurations preserving $W_p = +1$ for all plaquettes. Any spin configuration on a plaquette which violates this condition defines an excitation and is called a vortex, borrowing nomenclature from $Z_2$ gauge theory to which the model, in this limit, can be mapped. By simultaneously applying a pair of spin operators (Figs. 1**b** and 1**c**), one for each neighboring site, separated by a $z$-link and labeled as 1 and 2, we force $W_p = +1 \rightarrow -1$ on two neighboring plaquettes, thereby creating a pair of vortex excitations (vortices are always created in pairs). Here two spin operators are needed to preserve the alignment of the spins along the $z$-links, that is, the ground state subspace $\{|\uparrow\uparrow\rangle, |\downarrow\downarrow\rangle\}$. By definition, different types of vortices live on different sublattices of the honeycomb lattice and are called $e$ and $m$ - vortices [3]. Here sublattice means alternate rows of the lattice, and the choice of sublattice is irrelevant. These vortices (and combinations thereof) define the entire set of low energy excitations of the system.

We create vortices by applying the spin pair operation to site pairs which, in effect, rotates pairs of spins on $z$-links: $\exp(-i\bar{\sigma}^\alpha \tau)$, where an external field applied for a time $\tau$ reorients both spins. Such operators require control over single atoms at specific sites. However, it is not clear how one can apply a well controlled external potential to a single lattice site because the lattice spacing is on the order of the laser wavelength. Accordingly, systems at the diffraction limit will incorporate several sites at the same time and therefore prevent manipulation of spins of single atoms. A recent proposal [27] establishes a simple and efficient technique for selectively manipulating spin states of single atoms using a combination of focused lasers and microwave pulses, which, as we will show, enables the creation and manipulation of vortices through individual spin operations.

As a precursor to applying single spin operations, we first adiabatically ramp up the lattice. Adiabatic ramping of the lattice depth imposes a key simplification used prior to (and after) a set of single particle operations. Consider an adiabatic ramping up of the spin-independent optical lattice from an initial barrier $14E_R$ to around $25E_R$, while adjusting the spin-dependent lattice potential simultaneously so that the relation $J_z = 3J_x = 3J_y$ remains unchanged during the process. In the new lattice potential, the ground state wavefunction $|\psi_g\rangle$ does not change, while the time scale for spin-spin interactions is lengthened ($\hbar J_{\text{eff}}^{-1} \gg 10$s ), which means that changes in spin-spin interactions during local, fast operations (< 1ms) can be neglected. As a consequence, a series of single atom operations defining our braiding procedure act instantaneously (relative to $\hbar J_{\text{eff}}^{-1}$) on the highly correlated ground state. After spin operations on the ground state are completed we adiabatically lower the optical lattice potential depth. Note that we adiabatically ramp up (or down) the lattices in such a way that it merely decreases (or increases) the overall interaction energy scale, and does not perturb the structure of the spin Hamiltonian. Therefore such processes keep the state as the eigenstate of the Hamiltonian and the adiabatic time scale is limited by excitations to higher bands of lattices, instead of the spin-spin interaction strength.

We now discuss a scheme designed to implement a set of single spin operations after the adiabatic ramp up. In Fig. 2, we plot the atomic potential in a two dimensional, color-scale plot in the presence of a focused laser extending perpendicular to the honeycomb plane with an intensity maximum at a specific lattice site. The spatial distribution of the focused laser intensity induces position-dependent splittings between spin states $|\uparrow\rangle$ and $|\downarrow\rangle$. For the target atom, the focused laser ($\sigma^+$-polarized) induces a red-detuned trap for the spin state $|\downarrow\rangle$ with a depth chosen to be $V_\downarrow = -35 E_R$, but a blue-detuned trap for spin state $|\uparrow\rangle$ with depth $V_\uparrow = 18 E_R$ (corresponding to a power of 8.5μW with beam waist ~ 0.5μm). The wavelength of the laser is chosen to be $\lambda \approx 421 nm$, which corresponds to a detuning $\Delta_1 = -2\pi \times 1209 GHz$ from the transition $5^2 S_{1/2} \rightarrow 6^2 P_{3/2}$ to obtain the maximum ratio between the hyperfine splittings of two spin states and the spontaneous scattering rate [27].

A microwave pulse applied to the whole system will rotate the spin state of the target atom. The microwave frequency is chosen to be resonant with the hyperfine splitting of the target atom where the focused laser is applied, but has a detuning estimated to be $\hbar\delta \approx 52 E_R$ for non-target atoms. Different spin rotations $\sigma^x$ and $\sigma^y$ (note that $\sigma^z = i\sigma^x\sigma^y$ is a combination of $\sigma^x$ and $\sigma^y$), may be implemented using different phases $\varphi = \pi/2$ and $\varphi = 0$ of the microwave pulse, where $\varphi$ is defined through the magnetic field of the microwave $B \propto \cos(\vec{k} \cdot \vec{r} - \omega t + \varphi)$ with $\vec{r} = 0$ as the position of the target atom. A Gaussian shaped pulse $\Omega(t) = \Omega_0 \exp(-\omega_0^2 t^2)$ ($-t_f \leq t \leq t_f$) with parameters $\omega_0 = \delta/4$ and $\omega_0 t_f = 7$ (the pulse period $2t_f = 55\mu s$) is used to perform single spin operations. The variations of probabilities of non-target atoms in hyperfine states $|\uparrow\rangle$ and $|\downarrow\rangle$ caused by the microwave pulse are found from the Rabi equation to be smaller than $10^{-2}$. In addition, refocusing microwave pulses can be used to eliminate the phase variations of neighboring atoms due to the Rabi pulses [27]. By combining estimates from the adiabaticity criteria and the Rabi equation we find that the single spin operations may be accomplished in roughly 200μs (including ramping up and down of the focused laser, the microwave pulse period) and the probability to spontaneously scatter an unwanted photon is estimated to be small, $1.5 \times 10^{-4}$. The total probability for scattering a photon due to the focused laser is around $1 \times 10^{-2}$ in the whole braiding process which consists of about 60 single spin operations for the detection of anyonic statistics. In addition, the spontaneous emission probability due to the spin-dependent lattices is about $2 \times 10^{-2}$. The focused lasers need to be spatially stabilized because a displacement of the laser center from the minimum of the optical lattice potential induces a detuning of the microwave from the hyperfine splitting between two spin states of the target atom, and thus reduces the fidelity of the single spin rotation. For a small displacement 10nm, we estimate the detuning to be about $2\pi \times 350$Hz and find through integrating the Rabi equation that the fidelity of the rotation is degraded by $3 \times 10^{-3}$.

During the single spin operations (but after the adiabatic ramp up), we keep the lattice depth high which aids in defining multi-site operations. The single spin operations can be accomplished very fast ($\sim 0.2$ms) compared to $\hbar J_{eff}^{-1}$ and the double spin operation $\bar{\sigma}^y$ may be taken to be two consecutive single spin operations. Although each single operation $\sigma^y$ or $\sigma^x$ does not preserve the spin subspace $\{|\uparrow\uparrow\rangle, |\downarrow\downarrow\rangle\}$, the spin-spin interactions along the z-link which preserve the spin alignment are weak (almost zero) and can be neglected therefore two consecutive spin operations are equivalent to a double spin operation. This procedure allows for the creation and braiding of vortices at specifically chosen locations. Note that errors in braiding operations originating from imperfect single or double spin operations as well as the impact on non-target atoms can be automatically corrected by the topological properties of the Hamiltonian. When the optical lattice depth is lowered, the Kitaev Hamiltonian energetically penalizes unwanted local excitations. The result is a decay to the prepared topologically protected sector in the presence of a bath. However, the leakage errors due to spontaneously scattered photons are not automatically corrected because atoms are scattered to other hyperfine states, and thus are out of topological protected subspace of states.

After creating excitations, we need a braiding procedure that contains a series of spin operations in order to observe the topological phase. The $e$ and $m$-vortices discussed here are anyons because the wavefunction acquires a minus sign upon a full braid of one flavor of vortex around the other flavor (braiding around a vortex of the same flavor does not produce a sign change). A braid along a path $C$ is defined through a contiguous string of spin rotations traversing the lattice[3]: $R_C = \prod_{k \in C} \exp(-i\tau \bar{\sigma}_k^{\alpha_k})$, where the direction of the spin operator, $\alpha_k$, is determined by the direction of the move. Note that each move progresses by creating two new vortex excitations on neighboring plaquettes, which annihilates the original vortex on one plaquette and subsequently creates a vortex on the neighboring plaquette. Such process must be much faster than the time scale $\hbar J_{\text{eff}}^{-1} \gg 10s$ set by the excitation energy gap, which is clearly satisfied in our scheme. Figs. 1**d** and 1**e** show two types of moves, horizontal and vertical, and the associated spin operators, which can be accomplished by applying suitable procedures for single spin manipulation described above.

Fig. 3 shows two examples of braiding procedures; one $e$-vortex looping around another, Fig. 3**a**, which produces no sign change and one $e$-vortex taken around an $m$-vortex, Fig. 3**b**, which does produce a sign change. This minus sign arises from the anti-commutation relation of spin $\sigma_D^y$ (from path $C_T$) and $\sigma_D^z$ (from path $C_B$) at the site labeled $D$ in Fig. 3**b**. Initially two pairs of $e$-vortices are created by applying spin operations $\bar{\sigma}^z$ (Fig. 3**a**) at lattice sites $A$ and $B$ respectively. The left vortex of the pair at B is moved to the center of the lattice along a path $C_B$ by a series of spin operations $R_{C_B}$. The left vortex of the pair at A is then braided around the central vortex along a path $C_T$. The central vortex can be moved back to the original site by applying $R_{C_B}^{-1}$. Now both pairs of vortices are back to original pair location, where they are fused to vacuum by $\bar{\sigma}^z$ at sites $A$ and $B$. Because the paths $C_B$ and $C_T$ do not intersect at any lattice site, $R_{C_B}$ and $R_{C_T}$ commute and the final wavefunction is $|\psi_f\rangle = \bar{\sigma}_A^z \bar{\sigma}_B^z R_{C_B}^{-1} R_{C_T} R_{C_B} \bar{\sigma}_B^z \bar{\sigma}_A^z |\psi_g\rangle = |\psi_g\rangle$. There is therefore no net gain in an overall minus sign in a braid of an $e$-vortex around another $e$-vortex. The situation is different when an $e$-vortex is braided around an $m$-vortex as shown in Fig. 3**b** where the same procedure as that in Fig. 3**a** has been applied. Here, however the paths $C_B$ and $C_T$, defined through a serie of spin operators, intersect at lattice site $D$, where spin operators $\sigma_D^y$ from $R_{C_T}$ and $\sigma_D^z$ from $R_{C_B}$ are both applied. Because of the anti-commutation relation of $\sigma_D^y$ and $\sigma_D^z$, a minus sign is obtained when we exchange $R_{C_T}$ and $R_{C_B}$, that is $R_{C_T} R_{C_B} = -R_{C_B} R_{C_T}$. Therefore the final wavefunction is $|\psi_f\rangle = -|\psi_g\rangle$. We find a net gain of an overall minus sign in a braid of an $e$-vortex around an $m$-vortex.

We arrive at an important aspect of quasiparticle braiding and related statistics. The defining moment in braiding occurs at the braid crossing point. The notion of braiding statistics is topologically robust because the closed loop may acquire small fluctuations in shape due to external localnoise, but, as long as it is a closed loop about one $m$-vortex, the special point $D$ remains somewhere on the lattice. The spin states at the point $D$ provide an observable quantity useful in detecting anyonic braiding statistics.

We propose an interference experiment to observe the change in sign brought about by the braiding procedure. Consider two cases: an $e$-vortex braided around nothing, i.e. the vacuum state, which, after a full braid, leads to the original ground state, $|\psi_g\rangle$, and an $e$-vortex braided around an $m$-vortex, which leads to $-|\psi_g\rangle$. Taken separately the overall sign in each case is not directly observable. We create a superposition of both scenarios by simultaneously braiding the $e$-vortex around both the vacuum and an $m$-vortex, Fig. 4**a**. We generate this superposition by separating two $m$-vortices along the horizontal path $C_H$ with a sequence of $\pi/2$ pulses using the operations: $R_{C_H} = \prod_{k \in C_H} \exp(-i\pi \bar{\sigma}_k^z / 2)$, which creates a superposition of both the $m$-vortex state and the vacuum by virtue of the relation: $\exp(-i\pi \bar{\sigma}^{\alpha_j} / 2) = (I - i\bar{\sigma}^{\alpha_j})/\sqrt{2}$. We emphasize here that a sequence of $\pi/2$ pulses along the path $C_H$ is necessary. If initially we create a superposition of vacuum and an $m$-vortex pair by one $\pi/2$ pulse and then try to braid one vortex through a series of $\pi$ pulses, one $m$-vortex, instead of the superposition of vacuum and one $m$-vortex will be moved to the center. That is because the braiding operator creates a new $m$-vortex pair from vacuum, which is then braided by the $\pi$ pulses. Braiding an $e$-vortex along the closed loop $C_L$ via: $R_{C_L} = \prod_{k \in C_L} \exp(-i\pi \bar{\sigma}_k^{\alpha_k})$ closes our interference braid. To eliminate auxiliary vortices produced by the $\pi/2$ pulses we, as a final step, apply a series of $-\pi/2$ pulses along $C_H$. In Fig. 4**b**, with no $m$-vortex inside $C_L$, the final wavefunction $|\psi_1\rangle = \bar{\sigma}^z R_{C_L} \bar{\sigma}^z |\psi_g\rangle = |\psi_g\rangle$, is the same as the initial state. While in Fig. 4**a** an *m*-vortex (in a superposition with the vacuum) is created with $\pi/2$ pulses, the final wavefunction is then $|\psi_2\rangle = \bar{\sigma}^z R_{C_H}^{-1} R_{C_L} R_{C_H} \bar{\sigma}^z |\psi_g\rangle$ and is quite different from the initial ground state. At the intersection site $D'$, the path $C_H$ contains the operation $e^{-i\pi\bar{\sigma}^z/2} = (I - i\bar{\sigma}^z)/\sqrt{2}$, while $C_L$ contains the operation $-i\bar{\sigma}_{D'}^y$ and the commutation of them yields

$$|\psi_2\rangle = \frac{1}{2}(I_{D'} + i\bar{\sigma}_{D'}^z)^2 R_{C_H}^{-1} R_{C_H} \bar{\sigma}^z R_{C_L} \bar{\sigma}^z |\psi_g\rangle = i\bar{\sigma}_{D'}^z |\psi_g\rangle,$$ showing a pair of *m*-vortices at the site $D'$ (Fig. 4**c**). Had the *m*-vortex never been placed at the center of the loop $C_L$, the interference experiment would produce no signature at the point $D'$ and the system would return to its ground state, $|\psi_g\rangle$, upon a full braid, (see Fig. 4**b**). Therefore, detecting a pair of *m*-vortices at the location $D'$ in the interference experiment would provide concrete evidence for anyonic statistics.

Detecting the presence of two adjacent vortices (Fig.4**c**) is tantamount to observing the local spin-spin correlators, $T_i^{\alpha\beta} = \langle \psi_i | \sigma_{D'}^\alpha \sigma_F^\beta | \psi_i \rangle$ of two atoms at $D'$ and its *z*-link neighbor, $F$, where $\alpha, \beta = x, y, z$, $i = 1, 2$. Note that given different final states, $|\psi_1\rangle = |\psi_g\rangle$ and $|\psi_2\rangle = i\bar{\sigma}_{D'}^z|\psi_g\rangle$, we find $T_2^{xx} = -T_1^{xx}$ and $T_2^{yx} = -T_1^{yx}$, that is, the spin correlators have different signs contingent upon the existence of a pair of vortices at two neighboring plaquettes around $D'$ (Fig.4**c**). In addition, $T_1^{xx}$ and $T_1^{yx}$ cannot be zero simultaneously for the highly entangled topologically ordered ground state, which can be written as $|\psi_g\rangle = \mu_\uparrow |\uparrow\uparrow\rangle_{D'F} |\phi_\uparrow\rangle + \mu_\downarrow |\downarrow\downarrow\rangle_{D'F} |\phi_\downarrow\rangle$. Here $\mu_\uparrow$ and $\mu_\downarrow$ are superposition coefficients, $\phi_\uparrow$ and $\phi_\downarrow$ are wavefunctions of atoms at all other sites and satisfy the normalization conditions $\langle \phi_\uparrow | \phi_\uparrow \rangle = \langle \phi_\downarrow | \phi_\downarrow \rangle = 1$. In the topologically ordered ground state, the spins at sites $D'$ and $F$ are highly entangled with other spins, therefore $\mu_\uparrow$ and $\mu_\downarrow$ are both non-zero. For $J_x = J_y = 0$, sites $D'$ and $F$ are decoupled from other sites and $|\phi_\uparrow\rangle = |\phi_\downarrow\rangle$. As $J_x$ and $J_y$ become non-zero, the overlap $\langle \phi_\uparrow | \phi_\downarrow \rangle$ starts to decrease from 1, but is not zero for small $J_x, J_y$. Substituting $|\psi_g\rangle$ into the two spin correlators, we find $T_1^{xx} = \langle \psi_g | \sigma_{D'}^x \sigma_F^x | \psi_g \rangle = \langle \phi_\uparrow | \phi_\downarrow \rangle \text{Re}(\mu_\uparrow^* \mu_\downarrow)$ and $T_1^{yx} = \langle \psi_g | \sigma_{D'}^y \sigma_F^x | \psi_g \rangle = \langle \phi_\uparrow | \phi_\downarrow \rangle \text{Im}(\mu_\uparrow^* \mu_\downarrow)$ respectively. Clearly, $T_1^{xx} = T_1^{yx} = 0$ means that either $\mu_\uparrow$ or $\mu_\downarrow$ must be zero, which is impossible for the topological ordered ground state.

We see that a measurement of the sign change in $T_i^{\alpha\beta}$ can distinguish the two states $|\psi_g\rangle$ and $i\bar{\sigma}_{D'}^z|\psi_g\rangle$. Unfortunately, local spin correlations can only be measured by local operations which distinguish themselves from conventional time of flight imaging methods that measure collective effects of the whole system [28]. Here we propose a scheme to detect local spin correlations using local operations, which essentially establishes a probe to detect the presence of individual vortex pairs. We first note that the spin correlator between atoms at two sites $D'$ and $F$ can be written as $T_i^{\alpha\beta} = Tr_{D'F}(\sigma_{D'}^\alpha \sigma_F^\beta \rho_{D'F})$, where $\rho_{D'F} = Tr|\psi_i\rangle\langle\psi_i|$ is the local reduced density matrix of sites $D'$ and $F$ obtained by tracing out all other sites. This means that the spin correlation functions can be measured by detecting atoms in different measurement bases. For example, we find the spin correlator $T_i^{xx}$ can be obtained by measuring the probabilities of observing atoms $D'$ and $F$ in the basis $\{|++\rangle, |+-\rangle, |-+\rangle, |--\rangle\}$ using the relation $T_i^{xx} = P_{|++\rangle} + P_{|--\rangle} - (P_{|+-\rangle} + P_{|-+\rangle})$, where $|\pm\rangle \equiv (|\downarrow\rangle \pm |\uparrow\rangle)/\sqrt{2}$.

The experimental scheme is plotted and described using four steps as shown in Fig. 5. (**a**) Using single spin operations with focused lasers and microwave pulses, we apply $\pi/2$ pulses sequentially to both atoms $D'$ and $F$ along the $\sigma^y$ spin axis to transfer atoms to the new basis. (**b**) In order to prevent fluorescence signal from non-target atoms during further detection processing we transfer all atoms in the state $|\downarrow\rangle = |F=1, m_F=-1\rangle$ to the state $|F=1, m_F=1\rangle$ by two $\pi$ microwave pulses, then all atoms at the state $|\uparrow\rangle = |F=2, m_F=-2\rangle$ are transferred to the state $|\downarrow\rangle$ by another $\pi$ microwave pulse. (**c**) With the assistance of focused lasers, we select only atoms at sites $D'$ and $F$ and transfer them from state $|\downarrow\rangle$ back to $|\uparrow\rangle$. (**d**) A detection laser that is resonant with $|\uparrow\rangle \to |3\rangle \equiv |5^2P_{3/2}: F=3, m_F=-3\rangle$ is applied to detect the probability of finding the atoms at $|\uparrow\rangle$ (corresponding to the basis state $|-\rangle$). The fluorescence signal (the number of scattered photons) has three quantized levels, which correspond to states $|++\rangle, |--\rangle, |+-\rangle$ (or $|-+\rangle$), respectively. Repeating the entire experiment many times yields the probabilities $P_{|++\rangle}, P_{|--\rangle}$ and $P_{|+-\rangle} + P_{|-+\rangle}$, and thus determines the spin correlator $T_i^{xx}$. Similarly, we can measure the spin correlation function $T_1^{yx}$

with different basis states $\{|\hat{\mp}+\rangle,|\hat{\mp}-\rangle,|\hat{=}+\rangle,|\hat{=}-\rangle\}$, where $|\hat{\pm}\rangle = (|\downarrow\rangle \pm i|\uparrow\rangle)/\sqrt{2}$ define a basis for atom $D'$. The only difference for above processes in measuring $T_1^{xx}$ and $T_1^{yx}$ is that the $\pi/2$ pulse on atom $D'$ in step (a) is along the $\sigma^x$ spin axis. In discussing these steps, we have applied a very general technique, a measure of the two-spin correlation function to reveal the presence of excitations of a braided state at the specific location, $D'$, and therefore anyonic statistics through the fluorescence of selected atoms.

We have shown how to create, braid, and detect Abelian anyons in a spin model defined on a honeycomb optical lattice. Our proposed observation of anyonic statistics utilizes two important precursors necessary for topological quantum computation: i) establishing the existence of a topological phase of matter, and ii) defining a braiding and readout procedure for executing suitably defined elementary gate operations with the goal of using topological excitations for quantum computation. Our braiding and detection techniques can also be used to generate different types of excitations useful in creating a set of topologically protected quantum gates using non-Abelian anyons, which may be found in the model discussed here but in a different parameter regime or in different models implemented with optical lattices.

**Acknowledgements** We thank Alexei Kitaev, Chetan Nayak, and Ian B. Spielman for helpful discussions. This work was supported by ARO-DTO, ARO-LPS, and NSF.

**Figure Captions**

Figure 1: **a**, Links *x*, *y*, and *z* on a honeycomb plaquette, *p*, with sites depicted by open and filled circles. **b (c)**, A horizontal (vertical) pair of *e*-vortices created by the application of the spin operator, $\bar{\sigma}_1^z = \sigma_1^z \otimes I_2$ ($\bar{\sigma}_1^y = \sigma_1^y \otimes \sigma_2^x$) to two sites along a *z*-link, where $I$ is the unit operator. **d (e)**, Horizontal (Vertical) move of an *e*-vortex by repeated applications of $\bar{\sigma}^z$ ($\bar{\sigma}^y$) operators.

Figure 2: The two-dimensional plane plots the color scaled potential seen by atoms sitting in the honeycomb lattice but in the presence of a focused laser. Dark blue indicates the potential minimum for the spin down hyperfine state while dark red indicates the maximum. A schematic of the focused laser extends out of the plane. Microwave pulses drive the transitions between two spin states (the inset), but only for an atom at the center of the focused beam. Atoms at sites away from the center experience a weak potential which keeps the hyperfine levels off resonance.

Figure 3: **a**, A braid of an *e*-vortex, along a path $C_T$ (blue dotted line) starting from the point *A*, around an *e*-vortex which started from the point *B* and moved along a path $C_B$ (red dotted line). The top *e*-vortex forms a closed loop through a series of elementary moves generated from spin operators. When the *e*-vortices return to their starting positions, the resulting state is the same as the starting ground state, $|\psi_g\rangle$. **b**, The same as the top panel but for an *e*-vortex braided around an *m*-vortex. Here, spin commutation relations at the point *D* yield a final state $-|\psi_g\rangle$ indicating anyonic statistics between *e* and *m*-vortices.

Figure 4: **a**, Schematic showing a closed loop braid of an *e*-vortex, $C_L$, denoted by a red dotted line. The *e*-vortex is taken around a superposition state of an *m*-vortex and the vacuum placed at the center of the loop by a series of half-spin rotations ($\pi/2$ pulses) along the horizontal, blue dotted line, $C_H$. The crossing point, $D'$, carries an observable signature of anyonic statistics, a pair of *m*-vortices. **b**, The same as **a** but with no central *m*-vortex. **c**, The pair of *m*-vortices created at the crossing point $D'$.

Figure 5: Series of experimental steps used to measure the spin-spin correlation function of two spins. *A* indicates either one of the two spins while *C* indicates all other spins in the lattice.

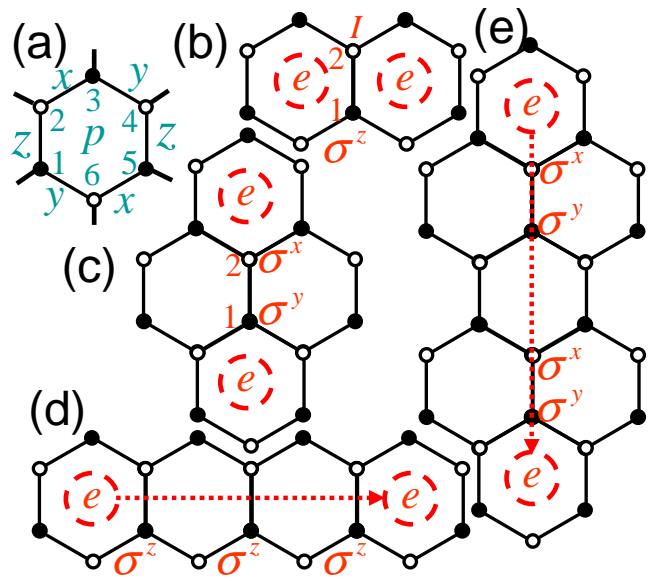

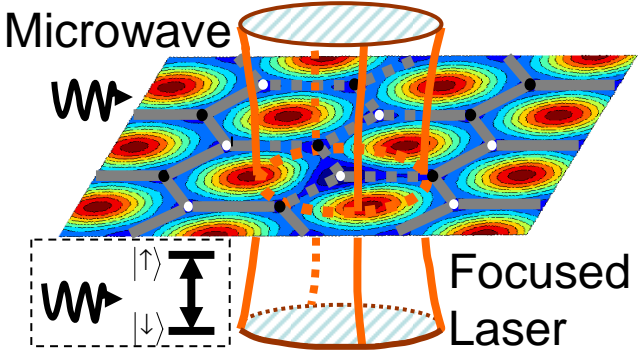

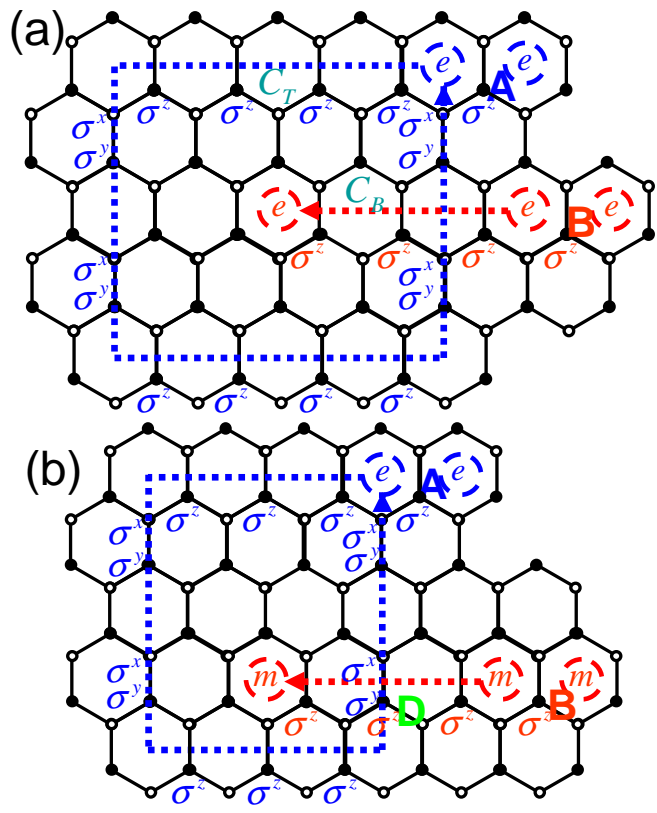

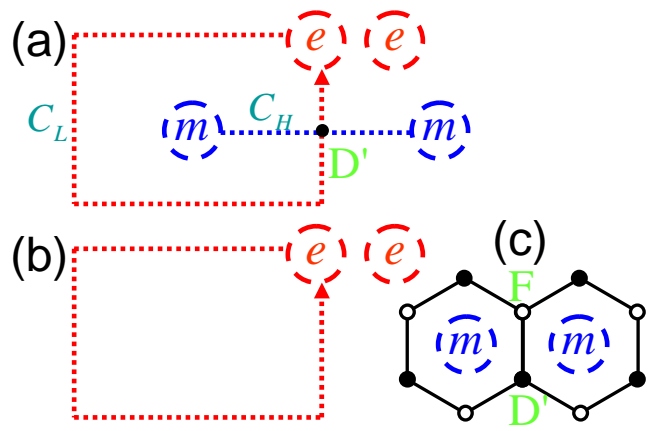

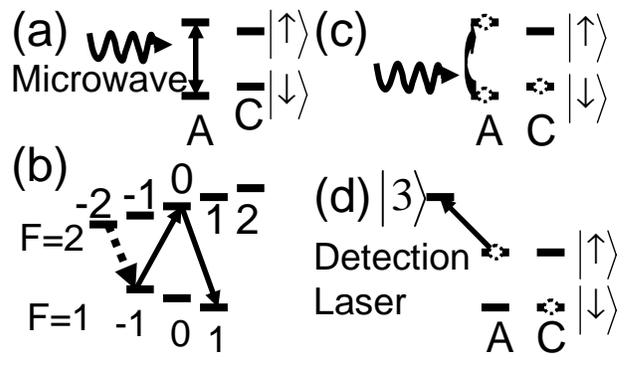